\documentstyle[12pt,epsf]{article}

\begin{document}

\font\fortssbx=cmssbx10 scaled \magstep2
\hbox to \hsize{
%\special{psfile=uwlogo.ps hscale=8000 vscale=8000 hoffset=-12 voffset=-2}
%\hskip.5in \raise.1in
\hbox{\fortssbx University of Wisconsin - Madison}
\hfill$\vcenter{\hbox{\bf MADPH-97-1007}
                \hbox{July 1997}}$ }

\vspace{1.5in}

\thispagestyle{empty}

\begin{center}
{\large\bf The AMANDA Neutrino Telescope:\\ Science Prospects and Performance at First Light}\\[3mm]
Francis Halzen
\end{center}

\vspace{1in}

\begin{abstract}
We update the science prospects for the recently completed AMANDA South Pole neutrino detector. With an effective telescope area of order $10^4$~m$^2$ and a threshold of $\sim$50~GeV, it represents the first instrument of a new generation of high energy neutrino detectors, envisaged over 25 years ago. We describe the instrument and its performance, and map its expansion to a detector of kilometer dimension.
\end{abstract}

\newpage
\section{A New Astronomy}

``And the estimate of the primary neutrino flux may be too low, since regions that produce neutrinos abundantly may not reveal themselves in the types of radiation yet detected" Greisen states in his 1960 review\cite{greisen}. He establishes that the natural scale of a deep underground neutrino detector is 15~m. This dream of neutrino astronomers is the same today. High energy neutrino telescopes are now multi-purpose instruments\cite{pr}; their science mission covers particle physics, astrophysics, cosmology and cosmic ray physics. Their deployment creates new opportunities for glaciology and oceanography, possibly geology of the Earth's core. The experimental techniques are, however, developed with the ultimate goal of deploying kilometer-size instruments. I will first introduce high energy neutrino detectors as astronomical telescopes using Fig.~1. The figure shows the diffuse flux of photons as a function of their energy and wavelength, from radio-waves to the high energy gamma rays detected with satellite-borne detectors\cite{turner}. The data span 19 decades in energy. Major discoveries have been historically associated with the introduction of techniques for exploring new wavelenghts. All of the discoveries were surprises; see Table~1. The primary motivation for commissioning neutrino telescopes is to cover the uncharted territory in Fig.~1: wavelengths smaller than $10^{-14}$~cm, or energies in excess of 10~GeV. This exploration has already been launched by truly pioneering observations using air Cerenkov telescopes\cite{weekes}. Larger space-based detectors and cosmic ray facilities with sensitivity above $10^7$~TeV, an energy where charged protons point back at their sources with minimal deflection by the galactic magnetic field, will be pursuing similar goals. Could the high energy skies in Fig.~1 be empty? No, cosmic rays with energies exceeding $10^8$~TeV have been recorded\cite{cronin}.

\begin{figure}[h]
\centering
\hspace{0in}\epsfxsize=3.5in\epsffile{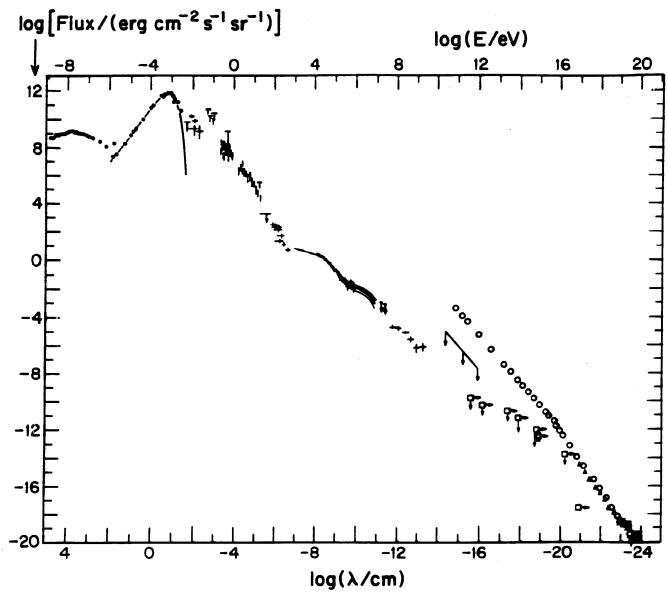}

\caption{Flux of gamma rays as a function of wavelength and photon energy. In the TeV--EeV energy range the anticipated fluxes are dwarfed by the cosmic ray flux which is also shown in the figure.}
\end{figure}

\begin{table}[t]
\caption{New windows on the Universe}
\bigskip
\centering
\begin{tabular}{|lcc|}
\hline
\quad\bf Telescope& \bf Intended use& \bf Actual results\\
\hline
optical (Galileo)& navigation& moons of Jupiter\\ 
radio (Jansky)& noise& radio galaxies\\
optical (Hubble)& nebulae& expanding Universe\\
microwave (Penzias-Wilson)& noise& 3K cosmic background\\
X-ray (Giacconi\dots)& moon& neutron stars\dots\\
radio (Hewish, Bell)& scintillations& pulsars\\
$\gamma$-ray (???)& thermonuclear explosions& $\gamma$-ray bursts\\
\hline
\end{tabular}
\end{table}

Astronomy with neutrinos does, however, have definite advantages. They can reach us, essentially without attenuation in flux, from the largest red-shifts. The sky is, in contrast, partially opaque to high energy photons and protons because of energy-loss suffered in interactions with infrared and CMBR photons\cite{cronin}. They do not reach us from distances much larger than tens of megaparsecs once their energy exceeds thresholds of 10~TeV for photons and $5\times 10^7$~TeV for protons. The drawback is that neutrinos are difficult to detect: the small interaction cross sections that enable them to travel without attenuation over a Hubble radius, are also the reason why kilometer-scale detectors are required in order to capture them in sufficient numbers to do astronomy\cite{halzenkm}. Some opportunities may, however, be unique. If, for instance, the sources of the highest energy cosmic rays are beyond $10^2$~Mpc, conventional astronomy is unlikely to discover them.

\section{Cosmic Beam Dumps}

Neutrinos are made in beam dumps. A beam of accelerated protons is dumped into a target where they produce pions in collisions with nuclei. Neutral pions decay into gamma rays and charged pions into muons and neutrinos. All this is standard particle physics and, in the end, roughly equal numbers of secondary gamma rays and neutrinos emerge from the dump. In man-made beam dumps the photons are absorbed in the dense target; this may not be the case in an astrophysical system where the target material can be more tenuous. Also, the most dense ``material" in the path of a cosmic beam may be light rather than nuclei. With an ambient photon density a million times larger than the sun, approximately $10^{14}$ per cm$^3$, particles accelerated in the superluminal jets associated with active galactic nuclei (AGN) may meet more photons than nuclei before losing energy. Examples of cosmic beam dumps are tabulated in Table~2. They fall into two categories. Neutrinos produced by the cosmic ray beam are, of course, guaranteed and calculable. We know the properties of the beam and the various targets: the atmosphere, the hydrogen in the galactic plane and the CMBR background. Neutrinos from AGN and GRB's (gamma ray bursts) are not guaranteed, though both represent good candidate sites for the acceleration of the highest energy cosmic rays. That they are also the sources of the highest energy photons reinforces this association.

\begin{table}
\caption{Cosmic Beam Dumps}
\bigskip
\centering
\tabcolsep=2em
\begin{tabular}{|c|c|}
\hline
\bf Beam& \bf Target\\
\hline
cosmic rays& atmosphere\\
cosmic rays& galactic disk\\
cosmic rays& CMBR\\
AGN jets& ambient light, UV\\
shocked protons& GRB photons\\
\hline
\end{tabular}
\end{table}

In order to accelerate a particle to energy $E$ in a magnetic field $B$, its gyroradius must be contained within the accelerator. In other words, its dimension $R$ has to exceed the gyroradius $E/B$. This leads to the relation
\begin{equation}
 E \leq BR,
\end{equation}
where the equality can be satisfied for a totally efficient accelerator. Applications of Eq.~1 are listed in Table~3. Its first entry illustrates the identification of supernova remnants with the acceleration of the bulk of the cosmic rays. The dimensional analysis of Eq.~1 yields an upper limit of $10^5$~TeV on the cosmic ray energy. Realistic modelling of supernova blast waves exploding into the interstellar medium, introduces inefficiencies which reduce this value by two orders of magnitude\cite{pr}, yielding a cutoff of $10^3$~TeV which we associate with the ``knee" in the cosmic ray spectrum. No galactic accelerator reaches $BR$-values of $10^8$~TeV; we therefore believe that the highest energy cosmic rays are accelerated outside our galaxy. There is experimental evidence to support this.

\begin{table}[h]
\caption{}
\bigskip
\centering

\hspace{0in}\epsfxsize=4in\epsffile{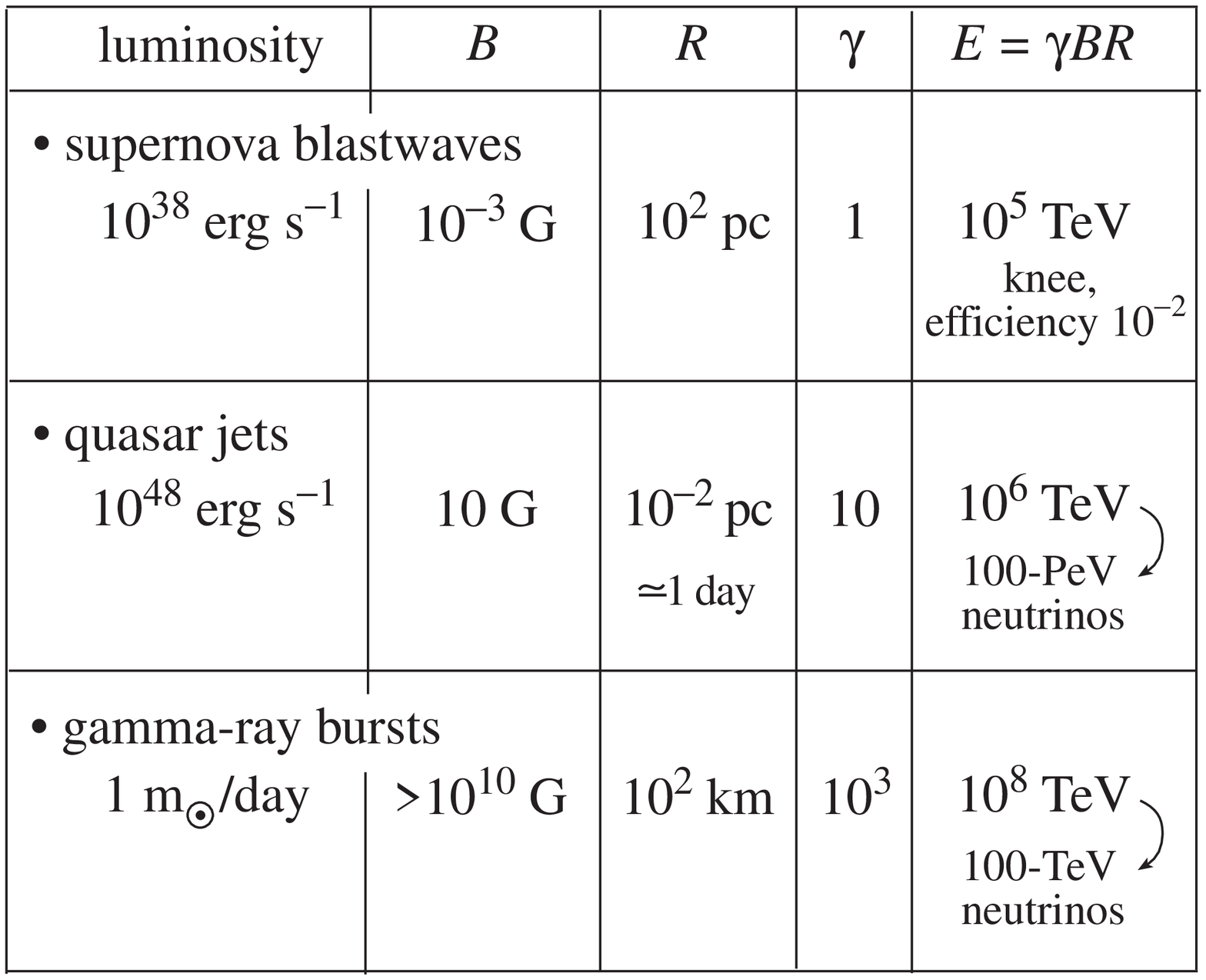}
\end{table}

The sources of the highest energy gamma rays, the jets associated with the supermassive black holes at the center of active galaxies and gamma ray bursts, are natural candidate sites for the acceleration of the highest energy cosmic rays; see Table~3. The jets in AGN consists of beams of electrons and protons accelerated by tapping the rotational energy of the black hole. The black hole is the source of the phenomenal AGN luminosity, emitted in a multi-wavelength spectrum extending from radio up to gamma rays with energies in excess of 20~TeV. In a jet BR-values in excess of $10^6$~TeV can be reached with fields of tens of Gauss extending over sheets of shocked material in the jet of dimension $10^{-2}$~parsecs; see Fig.~2. This size of the accelerating region is the source of the characteristic emission time, which is of order of a few days, sometimes less.  In an AGN jet Nature constructs a Fermilab accelerator about once a day! The $\gamma$-factor in the table reminds us that in the most spectacular sources the jet is beamed in our direction, thus increasing the energy and reducing the duration of the emission in the observer's frame. As previously mentioned, accelerated protons interacting with the ambient light are assumed to be the source of secondary pions which produce the observed gamma rays and, inevitably, neutrinos\cite{zas}.

\begin{figure}[h]
\centering
\hspace{0in}\epsfxsize=2.6in\epsffile{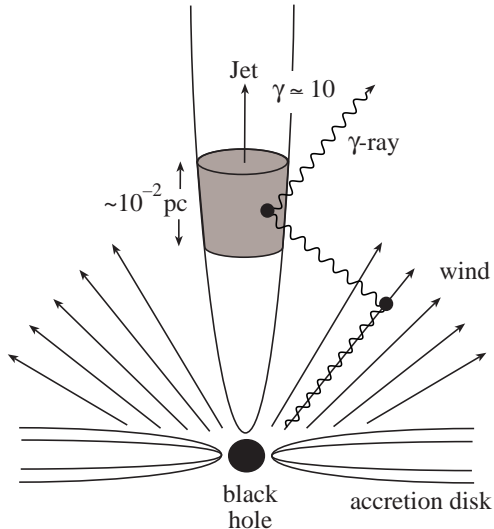}

\caption{Possible blueprint for the production of high energy photons and neutrinos near the super-massive black hole powering an AGN. Particles, accelerated in sheet-like bunches moving along the jet, interact with photons, radiated by the accretion disk or produced by the interaction of the accelerated particles with the magnetic field of the jet.}
\end{figure}

From our point of view, GRB's and AGN jets are similar. GRB's are somehow associated with the annihilation of neutron stars or solar mass black holes. Characteristic fields in excess of $10^{10}$~Gauss are concentrated in a fireball of size $10^2$~kilometers, which is opaque to light. The relativistic shock ($\gamma \simeq 10^3$) which dilutes the fireball to the point where the gamma ray display occurs, will also accelerate protons. These interact with the observed light to produce neutrinos\cite{waxman}; see Table~3.

Calculating the neutrino fluxes associated with the cosmic beam dumps listed in Table~3 is, at this point, straightforward\cite{pr}. The results are shown in Fig.~3a,b. We have displayed the neutrino flux multiplied by the square of their energy. Above 1~TeV the detection efficiency of neutrinos grows roughly linearly with energy, therefore the plot reflects the energy distribution of the observed events. In Fig.~3a we plot the guaranteed sources: atmospheric neutrinos, which are predominantly of charm origin above $10^3$~TeV, neutrinos emitted by the galaxy, in and out of the plane, and, finally, neutrinos produced by extra-galactic, $\sim 10^8$~TeV cosmic rays interacting with cosmic microwave photons. Assuming that cosmic ray acceleration is initiated in an earlier epoch, say at redshift $z=4$ rather than 2, increases the predicted flux. The fluxes from more speculative sources are shown in Fig.~3b: from the accretion flow and the jets of active galaxies, gamma ray bursts and topological defects; more about the latter later.

It is equally straightforward to compute the rate of detected muon-neutrinos in underground detectors\cite{pr}. The magnitude of the fluxes in Fig.~3 can be best gauged by the fact that the fluxes of AGN jets and of GRB's each yield roughly 100 neutrinos per year in a kilometer square telescope. The average neutrino energy from these sources are $10^2$ and $10^5$~TeV, respectively.

\renewcommand{\thefigure}{\arabic{figure}a}
\begin{figure}
\centering
\hspace{0in}\epsfxsize=3.3in\epsffile{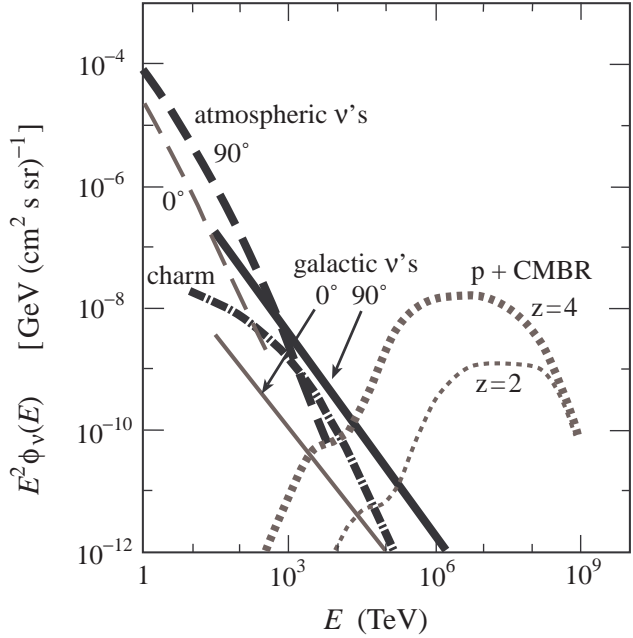}

\caption[]{Neutrino fluxes (from top to bottom, left to right): atmospheric neutrinos and prompt atmospheric neutrinos from charm decay for zenith angles of 90$^\circ$ and 0$^\circ$\cite{vazquez}; neutrinos produced by the interaction of cosmic rays with galactic matter for 0$^\circ$ and 90$^\circ$ galactic latitude\cite{domokos}; neutrinos produced by the interactions of extra-galactic cosmic rays with microwave photons for maximal redshift $z=4$ and $z=2$\cite{yoshida}.}
\end{figure}

\addtocounter{figure}{-1}\renewcommand{\thefigure}{\arabic{figure}b}
\begin{figure}[t]
\centering
\hspace{0in}\epsfxsize=3.3in\epsffile{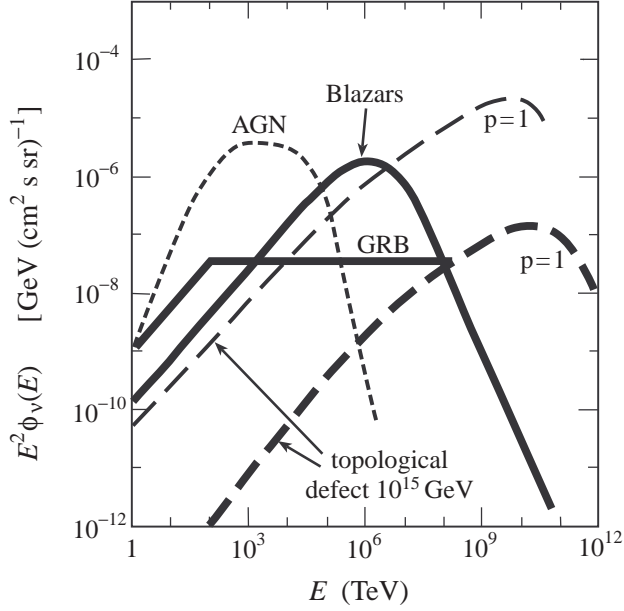}

\caption[]{Neutrino fluxes (from top to bottom, left to right): neutrinos produced in the accretion flow\cite{stecker} and jets\cite{zas} of AGN; neutrinos from GRB's\cite{waxman}; neutrinos from topological defects normalized to the flux of the highest energy cosmic rays assuming that they are protons (larger flux) and photons (lower flux)\cite{schramm}. The former should be correct because the highest energy cosmic rays are unlikely to be photons\cite{halzengamma}.}
\end{figure}
\renewcommand{\thefigure}{\arabic{figure}}

We would like to conclude this section with a comment on active galaxies. The typical  photon spectrum associated with an AGN jet is sketched in Fig.~4. The flux, from radio waves to X-rays, is produced by synchrotron radiation of the electron beam in the magnetic field of the jet. Notice however a second component of much higher energy in which, often, most of the luminosity of the galaxy is emitted. The conventional assumption has been that inverse Compton scattering by accelerated electrons generates these high energy photons. This has been challenged with models where accelerated protons, more efficient at tapping the energy of the black hole because of their reduced radiation via-a-vis electrons, produce pions in interactions with the plentiful ambient UV light. These are the parent particles of the high energy photons in Fig.~4. The observation of photons with energy in excess of 10~TeV can only be accounted for by fine-tuning of models without protons\cite{zas}. It is, however, more likely that only observation of neutrinos will conclusively settle this issue.

\begin{figure}[h]
\centering
\hspace{0in}\epsfxsize=3.8in\epsffile{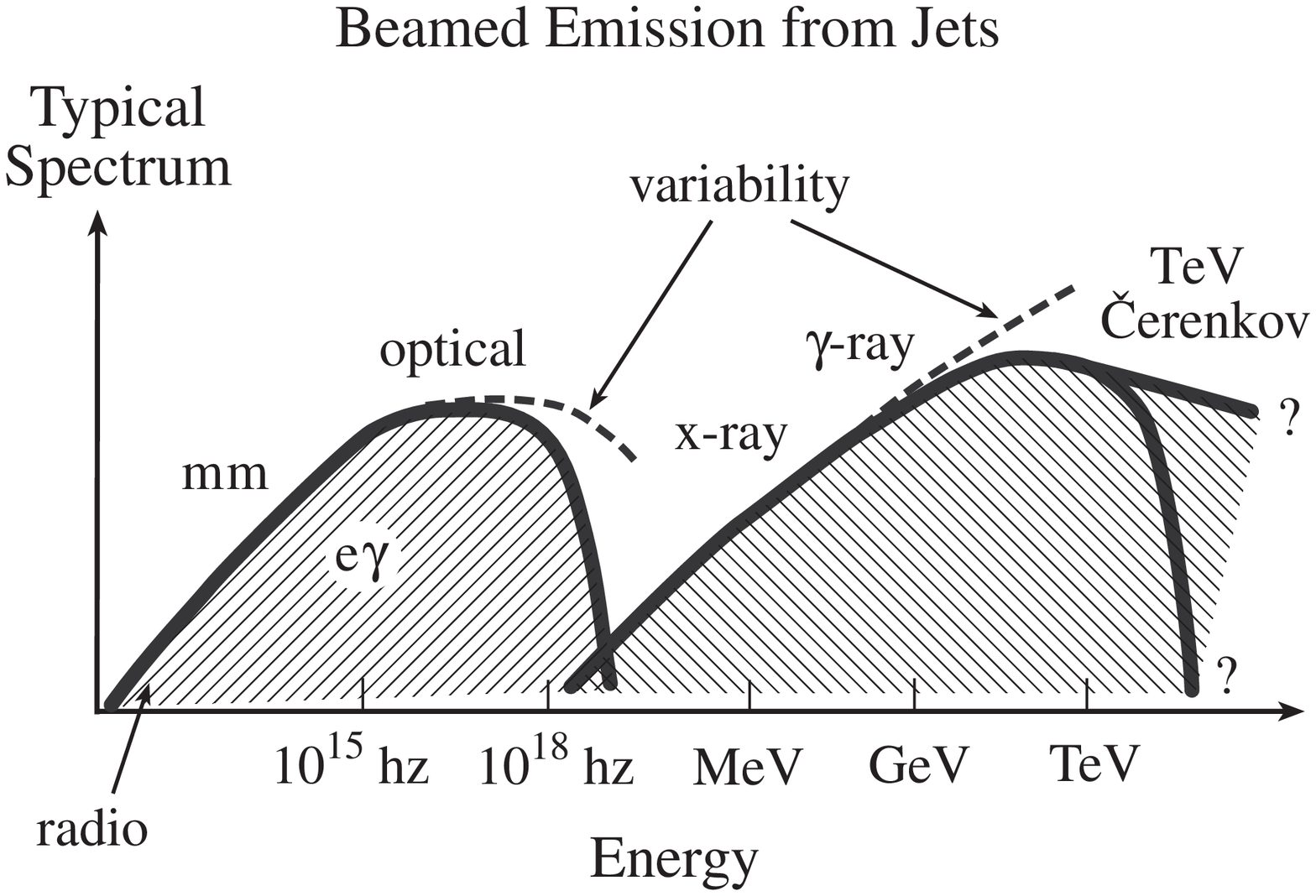}

\caption{Generic photon flux from the jet in an active galaxy.}
\end{figure}

\section{Particle Physics?}

If active galaxies turn out to be proton beam dumps, the intellectual connection of particle physics to astrophysics and cosmology will encompass standard observational astronomy. The intellectual connection of neutrino astronomy to particle physics is, however, more direct: neutrino mass, topological defects and supersymmetric dark matter are a few topics.

There are several opportunities for measuring neutrino mass:

$\bullet$
Detection of AGN or GRB's will allow us to study neutrinos over a baseline of  $10^3$~Mpc. Beam dumps produce electron- and muon-neutrinos. Therefore the appearance of tau-neutrinos in the detector is evidence for oscillations\cite{learnedpakvasa}. Above $5 \times 10^8$~TeV the travel path of a secondary tau exceeds the range of a muon. The detectors should therefore have excellent sensitivity to tau neutrinos.

$\bullet$
Ice telescopes are able to identify bursts of MeV-neutrinos in the sterile ice --- in water potassium decays swamp such a signal. In GRB observations one can clock the arrival time of neutrinos relative to photons over millisecond times and 1000~Mpc distance. Precision of $10^{-4}$~eV may be possible\cite{halzengrb}. This method for detecting a finite neutrino mass generated, arguably, the best present limit on the mass of the electron neutrino from observations of supernova 1987A.

$\bullet$
Finally, if oscillations are the origin of  the observation of a deficit of muon neutrinos in the atmospheric flux, the oscillation length is $\sim$200~km. Such oscillation will trace a single sinusoidal flux in a kilometer-deep detector with a threshold of tens of GeV\cite{blucher}; see Fig.~5. Confirmation of the so-called ``atmospheric anomaly" in a higher energy regime would undoubtedly be valuable.

\begin{figure}[h]
\centering
\hspace{0in}\epsfxsize=3.25in\epsffile{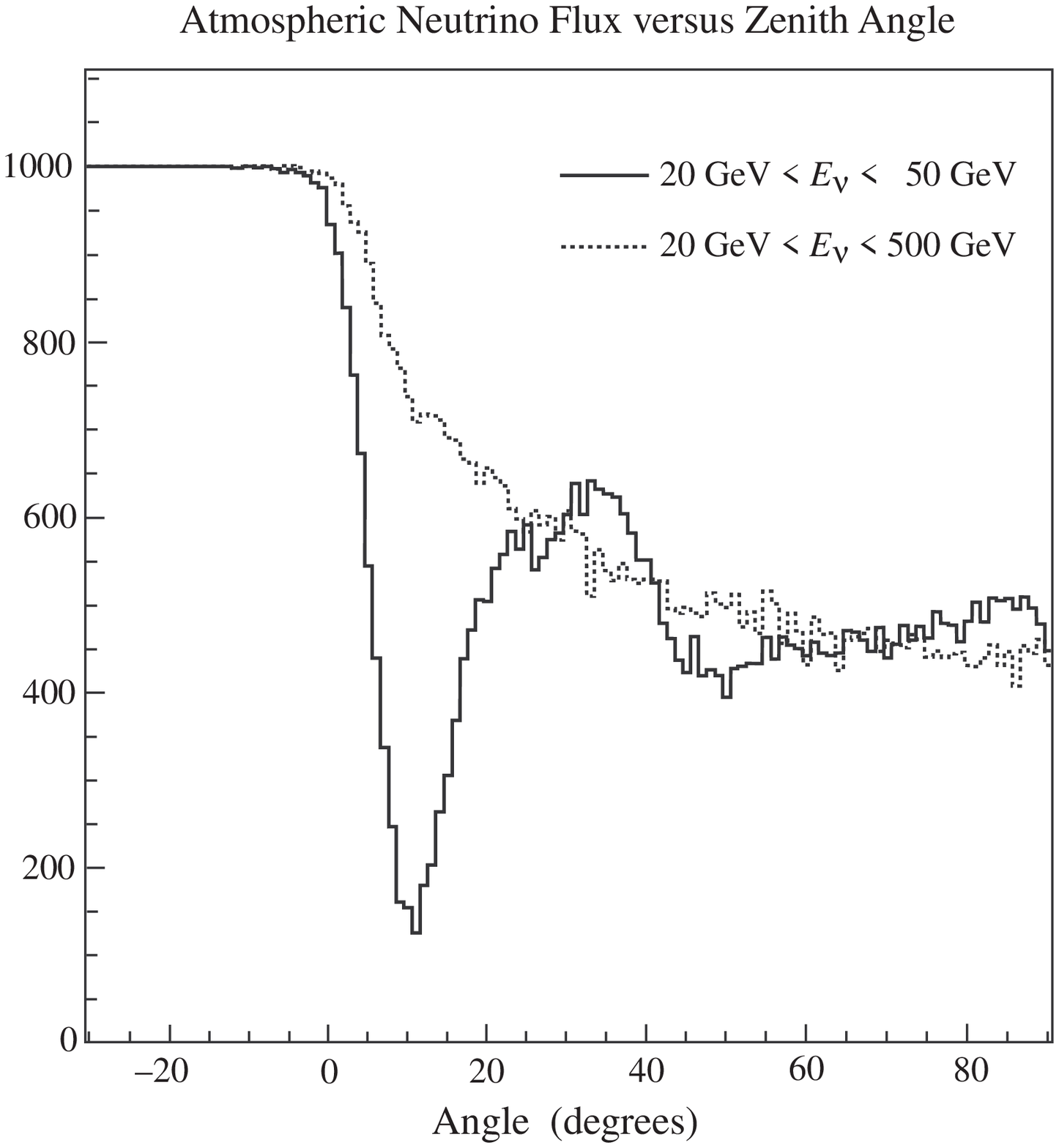}

\caption{Signature of oscillating atmospheric muon-neutrinos below the horizon (0--90$^\circ$) in the AMANDA detector.}
\end{figure}

If the stable, lightest supersymmetric particle impersonates the cold dark matter, neutrino telescopes of kilometer scale cover most of the parameter space of supersymmetric models\cite{kamionkowski}. We know that the cold dark matter particles have masses in excess of tens of GeV, otherwise they would have been detected at existing accelerators. Such heavy dark matter particles in our halo will, occasionally, annihilate into high energy neutrinos whose energy is one third of the mass, on average. For the highest masses, 500~GeV and above, neutrino telescopes become very effective at chasing supersymmetry. They are, in this sense, complementary to direct searches with LHC and Germanium-type detectors.

The ultimate particle physics within reach of neutrino telescopes covers topological defects. These structures represent an intrinsic feature of grand unified gauge theory which cannot be probed in accelerator experiments. Neutrinos constitute the dominant flux, by roughly two orders of magnitude, emitted by oscillating strings or annihilating monopoles\cite{schramm}.

\section{AMANDA: Towards ICE CUBE(D)}

\subsection{Status of the AMANDA Project}

First generation neutrino detectors are designed to reach a relatively large telescope area and detection volume for a neutrino threshold of tens of GeV, not higher. This relatively low threshold permits calibration of the novel instrument on the known flux of atmospheric neutrinos.  Its architecture is optimized for reconstructing the Cerenkov light front radiated by an up-going, neutrino-induced muon. Up-going muons are to be identified in a background of down-going, cosmic ray muons which are more than $10^5$ times more frequent for a depth of 1$\sim$2 kilometers. The method is sketched in Fig.~6.

\begin{figure}[h]
\centering
\hspace{0in}\epsfxsize=3.15in\epsffile{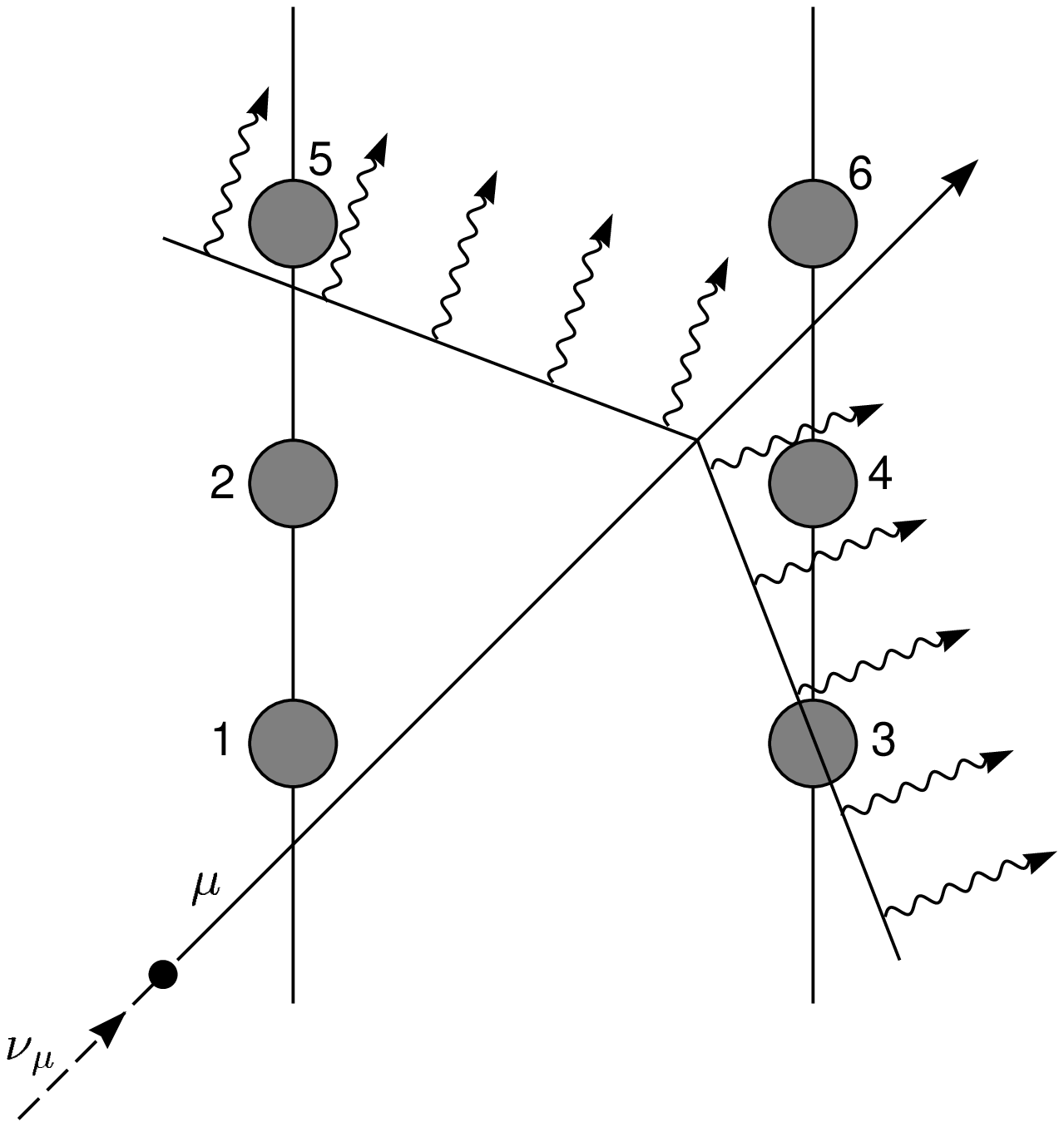}

\caption{The transit times of the Cerenkov photons in 6 optical modules determine the direction of the muon track.}
\end{figure}

Construction of the first-generation AMANDA detector\cite{barwick} was completed in the austral summer 96--97. It consists of 300 optical modules deployed at a depth of 1500--2000~m; see Fig.~7. An optical module (OM) consists of an 8~inch photomultiplier tube and nothing else. OM's have only failed when the ice refreezes, at a rate of less than 3 percent. Calibration of this detector is in progress, although data has been taken with 80 OM's which were deployed one year earlier in order to verify the optical properties of the ice (AMANDA-80).

\begin{figure}[h]
\centering
\hspace{0in}\epsfxsize=4.5in\epsffile{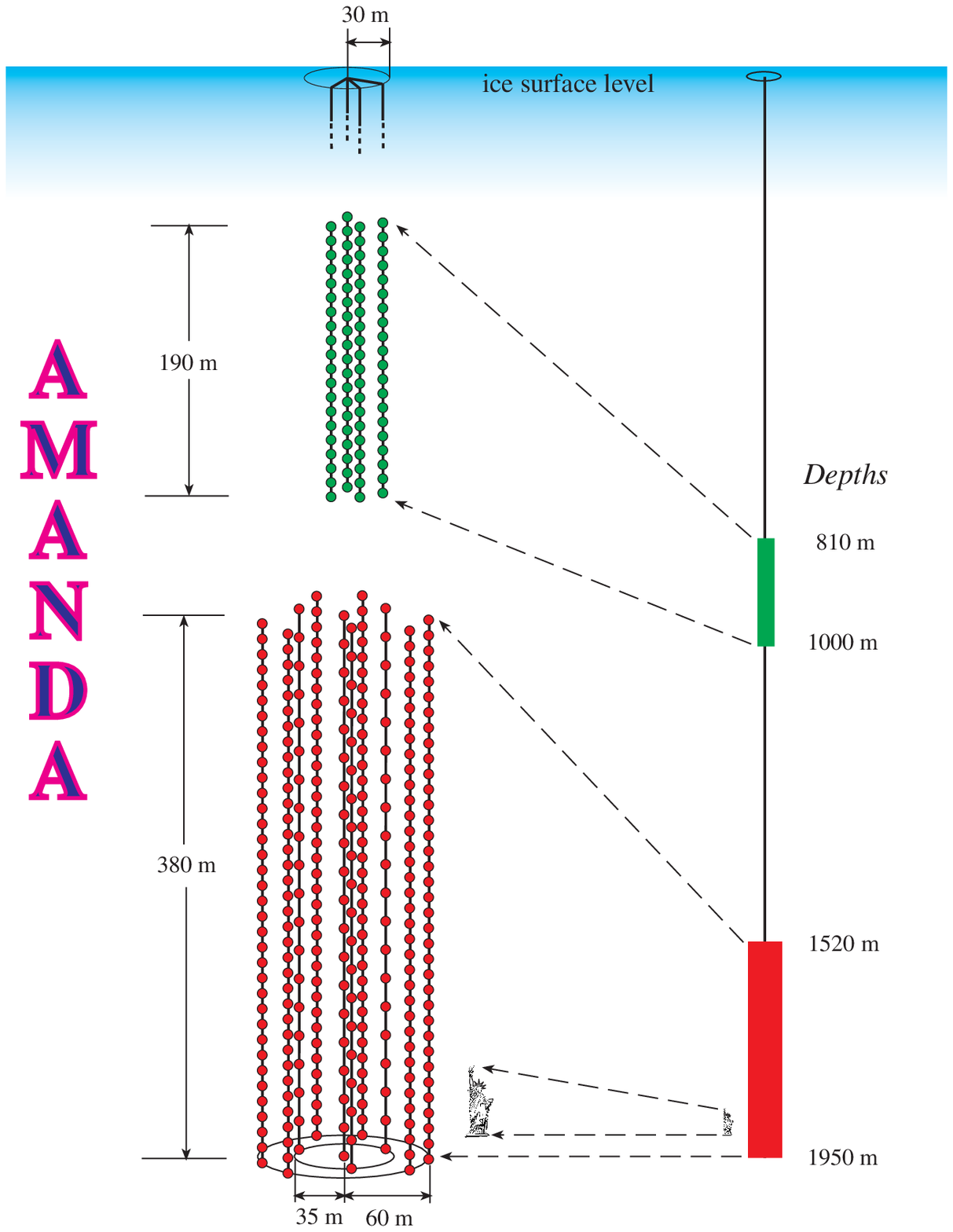}

\caption{}
\end{figure}

The performance of the AMANDA detector is encapsulated in the event shown in Fig.~8. Coincident events between AMANDA-80 and four shallow strings with 80 OM's (see Fig.~7), have been triggered for one year at a rate of 0.1~Hz. Every 10 seconds a cosmic ray muon is tracked over 1.2 kilometer. The contrast in detector response between the strings near 1 and 2~km depths is dramatic: while the Cerenkov photons diffuse on remnant bubbles in the shallow ice, a straight track with velocity $c$ is registered in the deeper ice. The optical quality of the deep ice can be assessed by viewing the OM signals from a single muon triggering 2 strings separated by 79.5~m; see Fig.~8b. The separation of the photons along the Cherenkov cone is well over 100~m, yet, despite some evidence of scattering, the speed-of-light propagation of the track can be readily identified.

\renewcommand{\thefigure}{\arabic{figure}a}
\begin{figure}[t]
\centering
\hspace{0in}\epsfysize=5.5in\epsffile{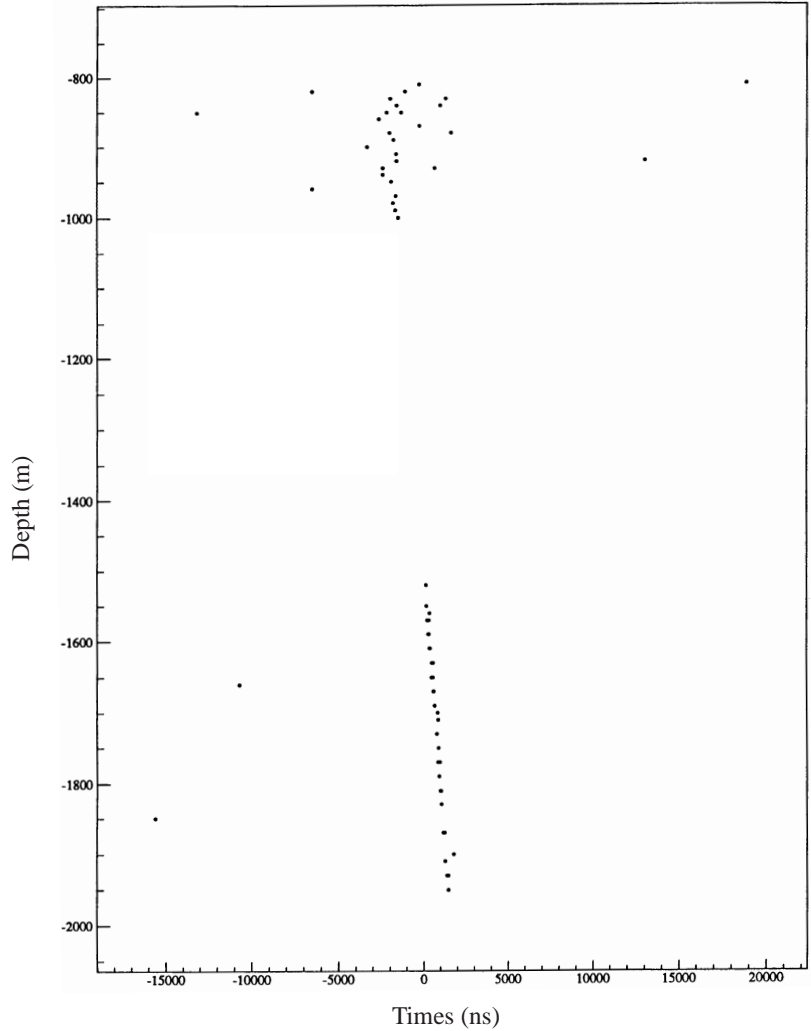}

\caption{Cosmic ray muon track triggered by both shallow and deep AMANDA OM's. Trigger times of the optical modules are shown as a function of depth. The diagram shows the diffusion of the track by bubbles above 1~km depth. Early and late hits, not associated with the track, are photomultiplier noise.}
\end{figure}

\addtocounter{figure}{-1}\renewcommand{\thefigure}{\arabic{figure}b}
\begin{figure}[t]
\centering
\hspace{0in}\epsfysize=5.5in\epsffile{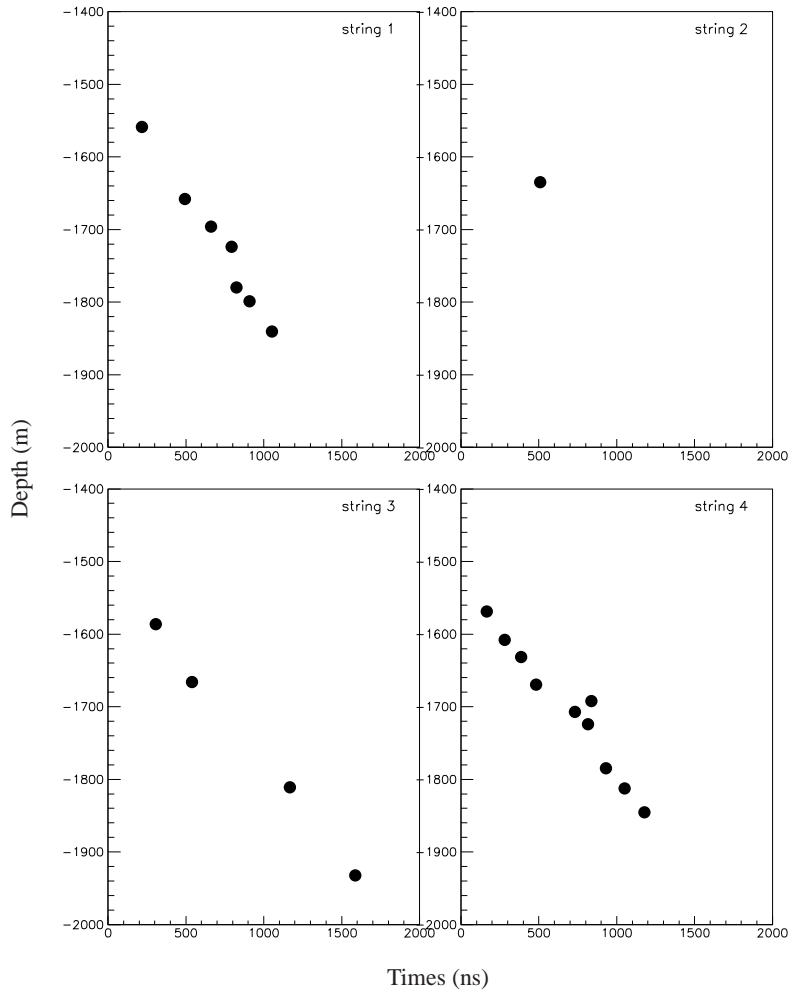}

\caption{Cosmic ray muon track triggered by both shallow and deep AMANDA OM's. Trigger times are shown separately for each string in the deep detector. In this event the muon mostly triggers OM's on strings 1 and 4 which are separated by 79.5~m. }
\end{figure}
\renewcommand{\thefigure}{\arabic{figure}}

The optical properties of the ice are quantified by studying the propagation in the ice of pulses of laser light of nanosecond duration. The arrival times of the photons after 20~m and 40~m are shown in Fig.~9 for the shallow and deep ice\cite{serap}. The distributions have been normalized to equal areas; in reality, the probability that a photon travels 70~m in the deep ice is ${\sim}10^7$ times larger. There is no diffusion resulting in loss of information on the geometry of the Cerenkov cone in the deep ice.

\begin{figure}[t]
\centering
\vspace{-1in}

\hspace{0in}\epsfxsize=5.5in\epsffile{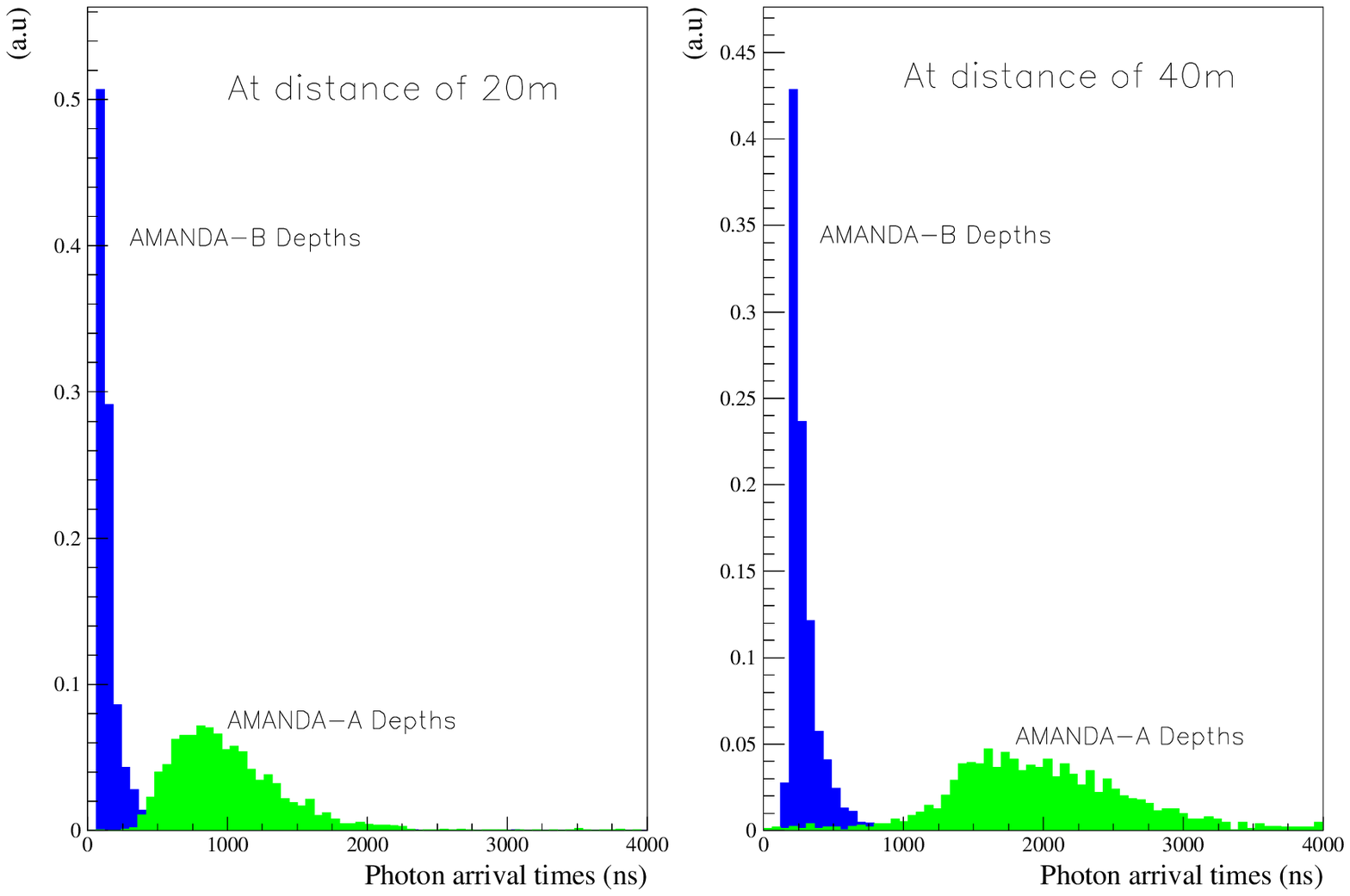}
\vspace{-2in}

\caption{Propagation of 510~nm photons indicate bubble-free ice below 1500~m, in contrast with ice with some remnant bubbles above 1~km.}
\end{figure}

\subsection{Intermezzo: AMANDA before and after}

The AMANDA detector was antecedently proposed on the premise that inferior properties of ice as a particle detector with respect to water could be compensated by additional optical modules. The technique was supposed to be a factor $5 {\sim} 10$ more cost-effective and, therefore, competitive. The design was based on then current information:

$\bullet$
the absorption length at 370~nm, the wavelength where photomultipliers are maximally efficient, had been measured to be 8~m,

$\bullet$
the scattering length was unknown,

$\bullet$
the AMANDA strategy was to use a large number of closely spaced OM's to overcome the short absorption length. Muon tracks triggering 6 or more OM's are reconstructed with degree accuracy. Taking data with a simple majority trigger of 6 OM's or more at 100~Hz yields an average effective area of $10^4$~m$^2$, somewhat smaller for atmospheric neutrinos and significantly larger for the high energy signals previously discussed.

The reality is that:

$\bullet$
the absorption length is 100~m or more, depending on depth\cite{science},

$\bullet$
the scattering length is $25 {\sim} 30$~m (preliminary),

$\bullet$
because of the large absorption length OM spacings are similar, actually larger, than that those of proposed water detectors. Also, a typical event triggers 20 OM's, not 6. Of these more than 5 photons are not scattered. In the end, reconstruction is therefore as before, although additional information can be extracted from scattered photons by minimizing a likelihood function which matches measured and expected delays\cite{christopher}.

\begin{figure}
\centering
\hspace{0in}\epsfxsize=5.5in\epsffile{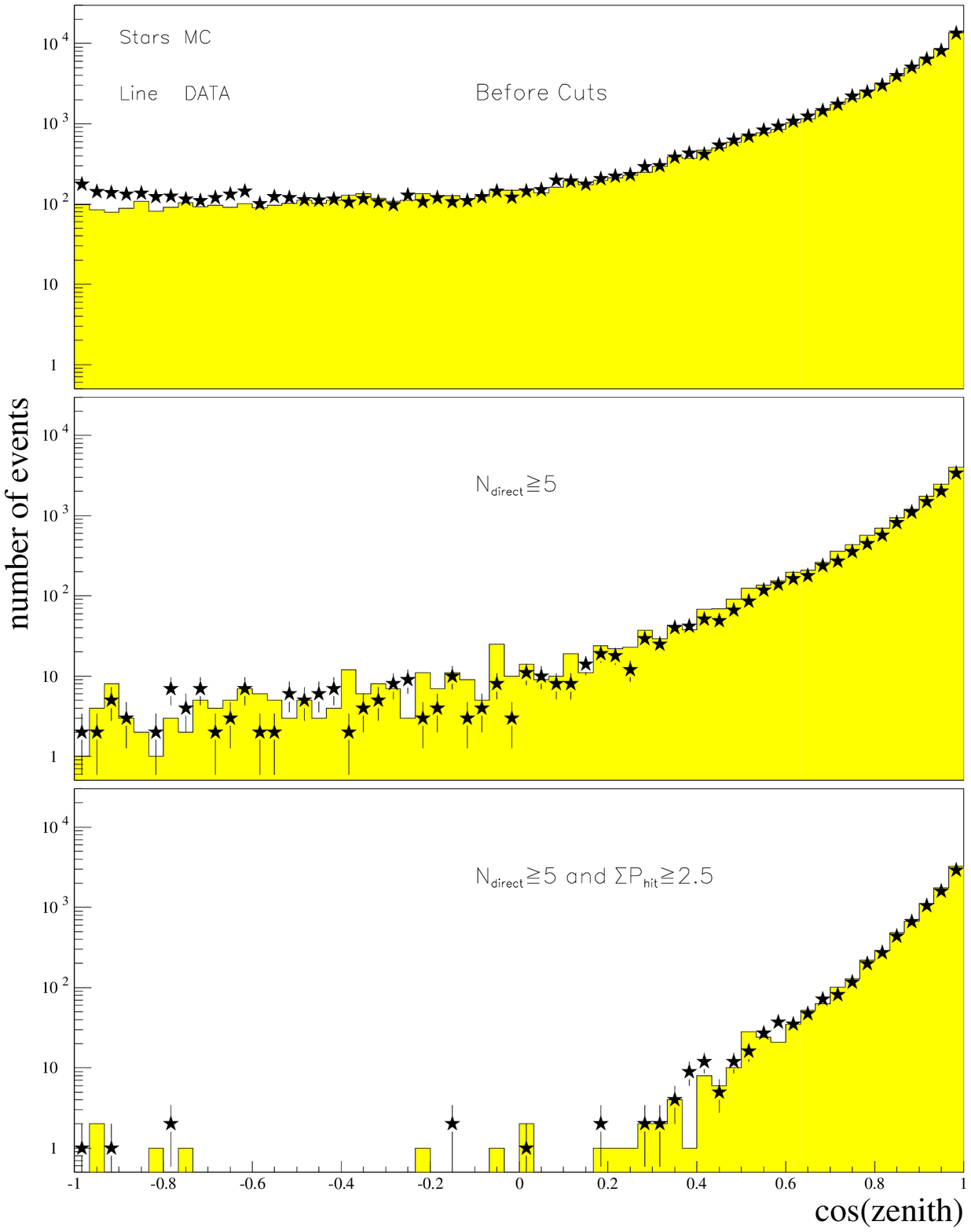}

\caption{Reconstructed zenith angle distribution of muons triggering AMANDA-80: data and Monte Carlo. The relative normalization has not been adjusted at any level. The plot demonstrates a rejection of cosmic ray muons at a level of 10~$^{-5}$.}
\end{figure}

The measured arrival directions of background cosmic ray muon tracks, reconstructed with 5 or more unscattered photons, are confronted with their known angular distribution in Fig.~10. The agreement with Monte Carlo simulation is adequate. Less than one in $10^5$ tracks is misreconstructed as originating below the detector\cite{serap}. Visual inspection reveals that the remaining misreconstructed tracks are mostly showers, radiated by muons or initiated by electron neutrinos, misreconstructed as up-going tracks of muon neutrino origin. At the $10^{-6}$ level of the background, candidate events can be identified; see Fig.~11.\footnote{There is an additional cut in Fig.~11 which simply requires that the track, reconstructed from timing information, actually traces the spatial positions of the OM's in the trigger. The power of this cut, especially for events distributed over only 4 strings, is very revealing. In a kilometer-scale detector, geometrical track reconstruction using only the positions of triggered OM's, is sufficient to achieve degree accuracy, as we will discuss further on.} This exercise establishes that AMANDA-80 can be operated as a neutrino detector; misreconstructed showers can be readily eliminated on the basis of the additional information on the amplitude of OM signals. Monte Carlo simulation, based on this exercise confirms, that AMANDA-300 is a $10^4$~m$^2$ detector with 2.5 degrees mean angular resolution\cite{christopher}. We have verified the angular resolution of AMANDA-80 by reconstructing muon tracks registered in coincidence with a surface air shower array\cite{miller}.

\begin{figure}[h]
\centering
\hspace{0in}\epsfxsize=3.4in\epsffile{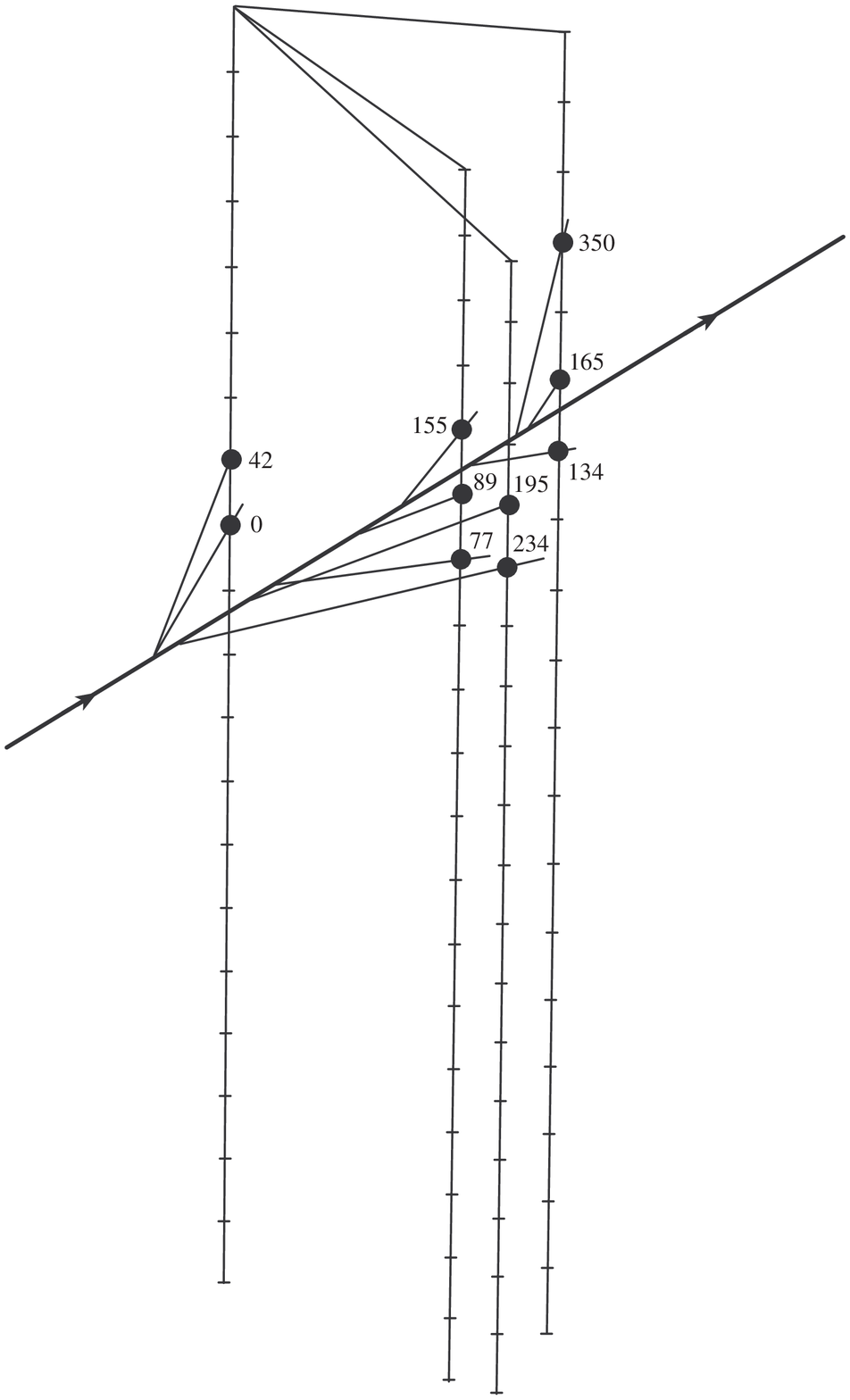}

\caption{A candidate up-going, neutrino-induced muon in the AMANDA-80 data. The numbers are times in nanoseconds, relative to the time of the first OM in the trigger.}
\end{figure}

\subsection{Towards ICE CUBE(D)}

A strawman detector with effective area in excess of 1~km$^2$ consists of 4800~OM's: 80 strings spaced by $\sim$~100~m, each instrumented with 60~OM's spaced by 15~m. A cube with a side of 0.8~km is thus instrumented and a through-going muon can be visualized by doubling the length of the lower track in Fig.~8a. It is straightforward to convince oneself that a muon of TeV energy and above, which generates single photoelectron signals within 50~m of the muon track, can be reconstructed by geometry only. The spatial positions of the triggered OM's allow a geometric track reconstruction with a precision in zenith angle of:
\begin{equation}
\mbox{angular resolution} \simeq \mbox{OM spacing/length of the track} \simeq \mbox{15\,m/800\,m} \simeq \mbox{1\,degree};
\end{equation}
no timing information is required. Timing is still necessary to establish whether a track is up- or down-going, not a challenge given that the transit time of the muon exceeds 2 microseconds.  Using the events shown in Fig.~8, we have, in fact, already demonstrated that we can reject background cosmic ray muons.

With half the number of OM's and half the price tag of the Superkamiokande and SNO solar neutrino detectors, the plan to commission such a detector over 5 years is not unrealistic. The price tag of the default technology used in AMANDA-300 is \$6000 per OM, including cables and DAQ electronics. We have also tested a new technology with very promising results. The charge of the photomultiplier is transformed into a light pulse by a LED\cite{albrecht}. This signal can be transmitted to the surface by fiber optic cable without loss of information. Given the scientific range and promise of such an instrument, a kilometer-scale neutrino detector must be one of the best motivated scientific endeavors ever.

\subsection{About Water\protect\cite{milla} and Ice}

The optical requirements of the detector medium can be readily evaluated, at least to first order, by noting that string spacings determine the cost of the detector. The attenuation length is the relevant quantity because it determines how far the light travels, irrespective of whether the photons are lost by scattering or absorption. Remember that, even in the absence of timing, hit geometry yields degree zenith angle resolution. Near the peak efficiency of the OM's the attenuation length is 25--30~m, slightly larger in ice at 1700~m than in water below 4~km. The advantage of ice is that, unlike for water, its transparency is not degraded for blue Cerenkov light of lower wavelength, a property we hope to take maximal advantage of by using wavelength-shifter in future deployments. Water is more transparent than ice for green light.

The AMANDA approach to neutrino astronomy was initially motivated by the low noise of sterile ice and the cost-effective detector technology. These advantages remain, even though we know now that water and ice are competitive as a detector medium. They are, in fact, complementary. Water and ice seem to have similar attenuation length, with the role of scattering and absorption reversed. As demonstrated with the shallow AMANDA strings\cite{porrata}, scattering can be exploited to range out the light and perform calorimetry of showers produced by electron-neutrinos and showering muons. Long scattering lengths in water may result in superior angular resolution, especially for the smaller, first-generation detectors. This can be exploited to reconstruct events far outside the detector in order to increase its effective volume.

\section*{Acknowledgements}

This research was supported in part by the U.S.~Department of Energy under Grant No.~DE-FG02-95ER40896 and in part by the University of Wisconsin Research Committee with funds granted by the Wisconsin Alumni Research Foundation.

\section*{The AMANDA Collaboration:}
P.~Askebjer$^1$, S.W.~Barwick$^2$, R.~Bay$^6$, R.~Van Berg$^9$, A.~Biron$^5$, L.~Bergstr\"om$^3$, A.~Bouchta$^1$, S.~Carius$^3$. E.~Dahlberg$^1$,  B.~Erlandsson$^1$, A.~Goobar$^1$, L.~Gray$^4$, A.~Hallgren$^3$, F.~Halzen$^4$, G.~Hill$^8$, P.O.~Hulth$^1$, S.~Hundertmark$^5$, J.~Jacobsen$^4$, S.~Johansson$^1$, V.~Kandhadai$^4$, A.~Karle$^4$, I.~Liubarsky$^4$, D.~Lowder$^6$,
T.C.~Miller$^7$, P.~Mock$^2$, R.~Morse$^4$, M.~Newcomer$^9$, D.~Nygren$^6$, R.~Porrata$^2$, P.B.~Price$^6$, A.~Richards$^6$, H.~Rubinstein$^3$, E.~Schneider$^2$, R.~Schwarz$^8$, C.~Spiering$^5$,
 O.~Streicher$^5$, Q.~Sun$^1$, T.~Thon$^5$, S.~Tilav$^4$, C.~Walck$^1$, C.~Wiebusch$^5$,
 R.~Wischnewski$^5$, K.~Woschnagg$^6$, G.~Yodh$^2$

\medskip\noindent
$^1$Dept.\ of Physics, Stockholm University, Sweden\\
$^2$Dept.\ of Physics, University of California, Irvine\\
$^3$Dept.\ of Physics, University of Uppsala, Sweden\\
$^4$Dept.\ of Physics, University of Wisconsin, Madison\\
$^5$DESY -- Inst.\ for High Energy Physics, Zeuthen, Germany\\
$^6$Dept.\ of Physics, University of California, Berkeley\\
$^7$Bartol Research Inst., University of Delaware, Newark\\
% $^8$Lawrence Berkeley Laboratory, California\\  OLD
$^8$NSF, Amundsen-Scott Station, South Pole\\
$^9$University of Pennsylvania, Philadelphia


\begin{thebibliography}{99}
\frenchspacing
\addtolength{\itemsep}{-.05in}
\bibitem{greisen}
K.~Greisen, {\it Ann. Rev. Nucl. Science}, {\bf 10}, 63 (1960); see also F.~Reines, {\it Ann. Rev. Nucl. Science}, {\bf 10}, 1 (1960); M.A. Markov \& I.M. Zheleznykh, {\it Nucl. Phys.} {\bf 27} (1961) 385; M.A. Markov in {\em Proceedings of the1960 Annual International Conference on High Energy Physics at Rochester}, E.~C.~G. Sudarshan, J.~H.~Tinlot \& A.~C.~Melissinos, Editors (1960).

\bibitem{pr}
For a review, see T.K. Gaisser, F. Halzen and T. Stanev, {\it Phys.\ Rep.} {\bf
258}(3), 173 (1995); R.~Gandhi, C.~Quigg, M.~H.~Reno and I.~Sarcevic, {\it Astropart. Phys.}, {\bf 5}, 81 (1996).

\bibitem{turner}
M.~T.~Ressel and M.~S.~Turner, {\it Comments Astrophys.} {\bf 14} (1990) 323.

\bibitem{weekes}
M.~Punch {\em et al.}, {\sl Nature} {\bf 358}, 477--478 (1992); J.~Quinn et
al., {\it Ap. J.} {\bf 456}, L83 (1995); Schubnell et al., astro-ph/9602068,  Ap. J. (1997 in press).

\bibitem{cronin}
The Pierre Auger Project Design Report, Fermilab report (Feb. 1997) and references therein.

\bibitem{halzenkm}
F. Halzen, {\it The case for a kilometer-scale neutrino detector}, in Nuclear and Particle Astrophysics and Cosmology, Proceedings of Snowmass~94, R.~Kolb and R.~Peccei, eds.

\bibitem{zas}
F.Halzen and E. Zas, astro ph/9702193, {\it Ap. J.}, in press (1997) and references therein.

\bibitem{waxman}
E.~Waxman and J.~N.~Bahcall, astro-ph/97011231, {\it Phys. Rev. Lett.} {\bf 78}, 2292 (1997).

\bibitem{vazquez}
E. Zas, F. Halzen and R. Vazquez, {\it Astropart. Phys.} {\bf1}, 297 (1993).

\bibitem{domokos}
G. Domokos {\it et al.}, {\it J. Phys. G} {\bf 19}, 899 (1993).

\bibitem{yoshida}
S. Yoshida and M. Teshima, {\it Prog. Theor. Phys.} {\bf 89}, 833 (1993).

\bibitem{stecker}
F. W. Stecker and M. Salomon, astro-ph/9501064.

\bibitem{halzengamma}
F. Halzen, R.A. Vazquez, T. Stanev and H.P. Vankov, {\it Astropart. Phys.} {\bf 3}, 151 (1995).

\bibitem{halzengrb}
F.~Halzen and G.~Jaczko, Phys. Rev. D {\bf 54}, 2774 (1996).

\bibitem{learnedpakvasa}
J. G. Learned and Sandip Pakvasa, {\it Astropart.\ Phys.} {\bf 3}, 267 (1995).

\bibitem{blucher}
E. Blucher, private communication.

\bibitem{kamionkowski}
G. Jungman, M. Kamionkowski amd K. Griest, {\it Phys. Rep.} {\bf 267}, 195 (1996).


\bibitem{schramm}
G. Sigl, D.N. Schramm and P. Bhattacharjee, astro-ph/9403039, {\it Astropart. Phys.} {\bf 2}, 401 (1994) and references therein.

\bibitem{barwick}
S. W. Barwick {\it et al.}, {\it The status of the AMANDA high-energy neutrino detector}, in Proceedings of the 25th International Cosmic Ray Conference, Durban, South Africa (1997).

\bibitem{serap}
S. Tilav {\it et al.}, {\it First look at AMANDA-B data}, in Proceedings of the 25th International Cosmic Ray Conference, Durban, South Africa (1997).

\bibitem{science}
The AMANDA collaboration, {\it Science} {\bf 267}, 1147 (1995).

\bibitem{christopher}
C. Wiebusch {\it et al.}, {\it Muon reconstruction with AMANDA-B}, in Proceedings of the 25th International Cosmic Ray Conference, Durban, South Africa (1997).

\bibitem{miller}
T. Miller {\it et al.}, {\it Analysis of SPASE-AMANDA coincidence events}, in Proceedings of the 25th International Cosmic Ray Conference, Durban, South Africa (1997).

\bibitem{albrecht}
A. Karle,  DESY Preprint 96-186, {\it First International Conference on New Developments in Photodetection}, to appear in {\it Proc. Suppl. Nucl. Inst. and Meth., Section A} (1997)

\bibitem{milla}
The status of the various experimental efforts can be found in {\it Proc.\ of the Seventh International Symposium on Neutrino Telescopes}, Venice (1996), ed.\ by  M.~Baldo-Ceolin, and in {\it Neutrino 96}, Helsinki 1996, World Scientific, M.~Roos, ed.

\bibitem{porrata}
R. Porrata {\it et al.}, {\it Analysis of cascades in AMANDA-A}, in Proceedings of the 25th International Cosmic Ray Conference, Durban, South Africa (1997).

\end{thebibliography}
\end{document}